\documentclass[12pt]{article}
\usepackage{amssymb}
\usepackage{epsfig}
\voffset= - 0.4in
\hoffset= - 0.6in         % switch off for draft style

\catcode`\@=11
\def\marginnote#1{}

\newcount\hour
\newcount\minute
\newtoks\amorpm
\hour=\time\divide\hour by60
\minute=\time{\multiply\hour by60 \global\advance\minute by-\hour}
\edef\standardtime{{\ifnum\hour<12 \global\amorpm={am}%
        \else\global\amorpm={pm}\advance\hour by-12 \fi
        \ifnum\hour=0 \hour=12 \fi
        \number\hour:\ifnum\minute<10 0\fi\number\minute\the\amorpm}}
\edef\militarytime{\number\hour:\ifnum\minute<10 0\fi\number\minute}

\def\draftlabel#1{{\@bsphack\if@filesw {\let\thepage\relax
   \xdef\@gtempa{\write\@auxout{\string
      \newlabel{#1}{{\@currentlabel}{\thepage}}}}}\@gtempa
   \if@nobreak \ifvmode\nobreak\fi\fi\fi\@esphack}
        \gdef\@eqnlabel{#1}}
\def\@eqnlabel{}
\def\@vacuum{}
\def\draftmarginnote#1{\marginpar{\raggedright\scriptsize\tt#1}}

\def\draft{\oddsidemargin -.5truein
        \def\@oddfoot{\sl preliminary draft \hfil
        \rm\thepage\hfil\sl\today\quad\militarytime}
        \let\@evenfoot\@oddfoot \overfullrule 3pt
        \let\label=\draftlabel
        \let\marginnote=\draftmarginnote
   \def\@eqnnum{(\theequation)\rlap{\kern\marginparsep\tt\@eqnlabel}%
\global\let\@eqnlabel\@vacuum}  }
\def\d{\partial}

\def\bea{\begin{eqnarray}}
\def\eea{\end{eqnarray}}

\def\beq{\begin{equation}}
\def\eeq{\end{equation}}
\def\ba{\beq\new\begin{array}{c}}
\def\ea{\end{array}\eeq}
\def\be{\ba}
\def\ee{\ea}
\def\stackreb#1#2{\mathrel{\mathop{#2}\limits_{#1}}}

\makeatletter
\newdimen\normalarrayskip              % skip between lines
\newdimen\minarrayskip                 % minimal skip between lines
\normalarrayskip\baselineskip
\minarrayskip\jot
\newif\ifold             \oldtrue            \def\new{\oldfalse}
\def\arraymode{\ifold\relax\else\displaystyle\fi} % mode of array entries
\def\eqnumphantom{\phantom{(\theequation)}}     % right phantom in eqnarray
\def\@arrayskip{\ifold\baselineskip\z@\lineskip\z@
     \else
     \baselineskip\minarrayskip\lineskip2\minarrayskip\fi}
\def\@arrayclassz{\ifcase \@lastchclass \@acolampacol \or
\@ampacol \or \or \or \@addamp \or
   \@acolampacol \or \@firstampfalse \@acol \fi
\edef\@preamble{\@preamble
  \ifcase \@chnum
     \hfil$\relax\arraymode\@sharp$\hfil
     \or $\relax\arraymode\@sharp$\hfil
     \or \hfil$\relax\arraymode\@sharp$\fi}}
\def\@array[#1]#2{\setbox\@arstrutbox=\hbox{\vrule
     height\arraystretch \ht\strutbox
     depth\arraystretch \dp\strutbox
     width\z@}\@mkpream{#2}\edef\@preamble{\halign
\noexpand\@halignto
\bgroup \tabskip\z@ \@arstrut \@preamble \tabskip\z@ \cr}%
\let\@startpbox\@@startpbox \let\@endpbox\@@endpbox
  \if #1t\vtop \else \if#1b\vbox \else \vcenter \fi\fi
  \bgroup \let\par\relax
  \let\@sharp##\let\protect\relax
  \@arrayskip\@preamble}
\def\eqnarray{\stepcounter{equation}%
              \let\@currentlabel=\theequation
              \global\@eqnswtrue
              \global\@eqcnt\z@
              \tabskip\@centering
              \let\\=\@eqncr
 \halign to \displaywidth\bgroup
    \eqnumphantom\@eqnsel\hskip\@centering
    $\displaystyle \tabskip\z@ {##}$%
    \global\@eqcnt\@ne \hskip 2\arraycolsep
         $\displaystyle\arraymode{##}$\hfil
    \global\@eqcnt\tw@ \hskip 2\arraycolsep
         $\displaystyle\tabskip\z@{##}$\hfil
         \tabskip\@centering
    &{##}\tabskip\z@\cr}
\begingroup\ifx\undefined\newsymbol \else\def\input#1 {\endgroup}\fi
\newfont{\hr}{msbm10}
\newfont{\ams}{msam10}

\font\numbers=cmss12
\font\upright=cmu10 scaled\magstep1
\def\stroke{\vrule height8pt width0.4pt depth-0.1pt}
\def\topfleck{\vrule height8pt width0.5pt depth-5.9pt}
\def\botfleck{\vrule height2pt width0.5pt depth0.1pt}
\def\Zmath{\vcenter{\hbox{\numbers\rlap{\rlap{Z}\kern 0.8pt\topfleck}\kern
2.2pt
                   \rlap Z\kern 6pt\botfleck\kern 1pt}}}
\def\Qmath{\vcenter{\hbox{\upright\rlap{\rlap{Q}\kern
                   3.8pt\stroke}\phantom{Q}}}}
\def\Nmath{\vcenter{\hbox{\upright\rlap{I}\kern 1.7pt N}}}
\def\Cmath{\vcenter{\hbox{\upright\rlap{\rlap{C}\kern
                   3.8pt\stroke}\phantom{C}}}}
\def\Rmath{\vcenter{\hbox{\upright\rlap{I}\kern 1.7pt R}}}
\def\Z{\ifmmode\Zmath\else$\Zmath$\fi}
\def\Q{\ifmmode\Qmath\else$\Qmath$\fi}
\def\N{\ifmmode\Nmath\else$\Nmath$\fi}
\def\C{\ifmmode\Cmath\else$\Cmath$\fi}
\def\R{\ifmmode\Rmath\else$\Rmath$\fi}
\def\stackreb#1#2{\mathrel{\mathop{#2}\limits_{#1}}}

\def\res{{\rm res}}
\def\Bf#1{\mbox{\boldmath $#1$}}
\def\balpha{{\Bf\alpha}}

\def\bmu{{\Bf\mu}}

\def\d{\partial}

\def\Im{{\rm Im}}
\def\Re{{\rm Re}}

\def\half{{\textstyle{1\over2}}}
\def\2{{1\over 2}}
\def\ntwo{${\mathcal N}=2\;$}

\def\none{${\mathcal N}=1\;$}

\def\beq{\begin{equation}}
\def\eeq{\end{equation}}
\def\ba{\beq\new\begin{array}{c}}
\def\ea{\end{array}\eeq}
\def\be{\ba}
\def\ee{\ea}
\def\stackreb#1#2{\mathrel{\mathop{#2}\limits_{#1}}}

\newcommand{\rf}[1]{(\ref{#1})}

\begin{document}

%\draft                               %SWITCH ON/OFF DRAFT VERSION%
\begin{flushright}
FIAN/TD-02/10\\
ITEP/TH-05/10
\end{flushright}

\vspace*{0.5cm}
%\vfil

\renewcommand{\thefootnote}{\fnsymbol{footnote}}
\begin{center}
\baselineskip20pt
{\bf \LARGE Period Integrals, Quantum Numbers\\
\vspace{0.3cm}
and Confinement in SUSY QCD\footnote{Contribution to special volume on {\em Integrable Systems in Quantum Theory }.}}
\end{center}
\vspace{1.0 cm}
\begin{center}
{\Large A.~Marshakov}\\
\vspace{0.6 cm}
{\em
Theory Department, P.N.Lebedev Physics Institute,\\
Institute of Theoretical and Experimental Physics,\\ Moscow, Russia
}\\
\vspace{0.3 cm}
{e-mail:\ \ mars@lpi.ru,\ \ mars@itep.ru}
\end{center}
\bigskip\bigskip\medskip

\begin{center}
\begin{quotation}
\noindent
We present a direct computation of the period integrals on degenerate Seiberg-Witten curves
for supersymmetric QCD, and show how these periods determine the changes in the quantum
numbers of the states, when passing from the weak to the strong-coupling domains in the mass moduli space of the theory. The confinement of monopoles at strong coupling is discussed, and we
demonstrate that the ambiguities in choosing the way in the moduli space do not influence
to the physical conclusions on confinement of monopoles in the phase with the condensed light dyons.
\end{quotation}
\end{center}

\renewcommand{\thefootnote}{\arabic{footnote}}
\setcounter{section}{0}
\setcounter{footnote}{0}
\setcounter{equation}0
\section{Introduction}

Supersymmetric QCD serves for a while as a laboratory for testing confinement. Not being
enough realistic to describe real nature, it can be considered
nevertheless as model for quantum theory, which allows non-perturbative analysis,
and therefore can prolong our horizon to understand, at least in principle, what happens with
gauge theory at strong coupling.

Below I am going to present some details of studying the properties of confinement
in supersymmetric QCD along the program of \cite{MY1,MY2}, and complete the discussion
of few technical issues, arising along these lines in underlying complex geometry. The main
idea of this scenario is to start with an obvious confinement of monopoles of the (dual) Meissner type at the quark vacuum in weakly coupled supersymmetric gauge theory \cite{MY1}, and then move this picture by adjusting the mass parameters of the theory towards the strongly coupled domain \cite{MY2}.
One ends up in this way with the effective theory of light dyons
instead of original quarks, since the BPS-states change their quantum numbers due to nontrivial monodromies in mass-moduli space.
Investigating of these monodromies is a nontrivial problem (see
e.g. \cite{BF,momon}) and can be performed in the most transparent way by computing the
period integrals and studying the perturbation of ramification points on (almost) singular
Seiberg-Witten curves \cite{SW1,SW2,HO,APS}.

\begin{figure}[tp]
\epsfysize=6cm
\centerline{\epsfbox{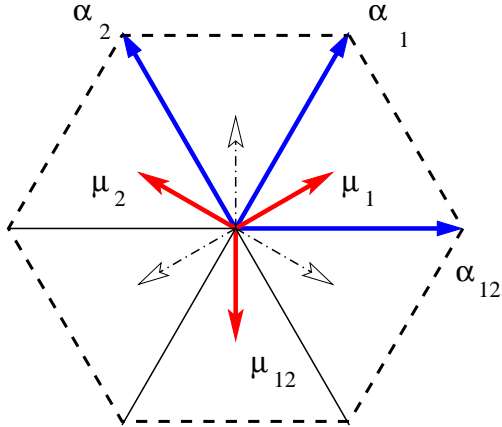}}
\caption{Roots $\balpha$ and fundamental weights $\bmu$ for the Lie algebra of the
$SU(3)$ gauge group in its Cartan plane, the roots are canonically normalized to
$\balpha^2=2$. The notations are chosen for the roots to be orthogonal
$\bmu_I\cdot\balpha_J=\delta_{IJ}$ ($I,J=1,2$) with the weights $\bmu_1$ and $\bmu_2$ of the
fundamental representations ${\bf 3}$ (the weights of the dual fundamental
representation ${\bar{\bf 3}}$ are depicted with dashed lines). Generic duality $\bmu\cdot\balpha\in\mathbb{Z}$ for the arbitrary vectors $\bmu$ from the
weight lattice and $\balpha$
(from its root sub-lattice) turn into the Dirac quantization condition for the (chromo) electric
and magnetic charges.}
\label{fi:su3}
\end{figure}

\begin{figure}[tp]
\epsfysize=3.5cm
\centerline{\epsfbox{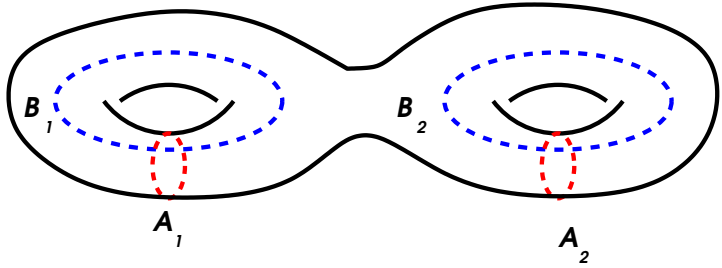}}
\caption{$A$- and $B$- cycles and their intersection form $A_i \circ B_j =\delta_{ij}$ for
the genus $N_c-1=2$ Riemann surface of the $SU(3)$ gauge theory. Elementary quark's charges $\bmu_{1,2}\leftrightarrow A_{1,2}$ correspond to the $A$-cycles, while the monopole's
ones are $\balpha_{1,2}\leftrightarrow B_{1,2}$ - to the $B$-cycles.}
\label{fi:riem2}
\end{figure}

Following \cite{MY1,MY2}, below the supersymmetric QCD with the $SU(N_c)$ gauge group and large number $N_c\leq N_f \leq 2N_c$ of the fundamental flavors is taken as a basic model: the
first nontrivial (and the main in this text) example is the $SU(3)$ gauge theories with
$N_f=4$ and $N_f=5$. The
quantum numbers of the light states can be seen on fig.~\ref{fi:su3}, where the
quark color charges (the fundamental weights) and monopole charges (the roots) for the
$SU(3)$ gauge group are depicted.
%%%%%%%%%%%%%%%%%%%%%%%%%%%%%%%%%%%%%%%%%%%%%%%%%%%%%%%%%%%%%%%%%%%%%%%%%%%%%%%%%%%%%%%%%%%%%%%%
%%%%%%%%%%%%%%%%%%%%%%%%%%%%%%%%%%%%%%%%%%%%%%%%%%%%%%%%%%%%%%%%%%%%%%%%%%%%%%%%%%%%%%%%%%%%%%%%%
In the context of Seiberg-Witten theory \cite{SW1,SW2}, which is necessary to
study the {\em exact}
properties of supersymmetric QCD around strongly-coupled vacua, the Dirac quantization condition $\bmu_i\cdot\balpha_j=\delta_{ij}$ turns into the intersection form of the
cycles $A_i \circ B_j =\delta_{ij}$ on spectral curve, see fig.~\ref{fi:riem2}
for the $SU(3)$ gauge group.
In what follows we shall use only the ``homological normalization'' of the charges \cite{SW2},
where they are measured by the cycles on Seiberg-Witten curve, and therefore are always {\em integer}.
%%%%%%%%%%%%%%%%%%%%%%%%%%%%%%%%%%%%%%%%%%%%%%%%%%%%%%%%%%%%%%%%%%%%%%%%%%%%%%%%%%%%%%%%%%%%%%%%
%%%%%%%%%%%%%%%%%%%%%%%%%%%%%%%%%%%%%%%%%%%%%%%%%%%%%%%%%%%%%%%%%%%%%%%%%%%%%%%%%%%%%%%%%%%%%%%%%

Duality transformations do not change the complex structure on the Seiberg-Witten curves,
but exchanges electric $A$-cycles with the magnetic $B$-cycles, and hence correspond to electric-magnetic duality \cite{SW1,SW2}. It is important, that the periods
\be
\label{SWper}
a_i={1\over 2\pi i}\oint_{A_i}dS,\ \ \ \
a^D_i={1\over 2\pi i}\oint_{B_i}dS
\ee
entering (together with the residues $m_A = \res_{P_A} dS$) the BPS mass-formula
\be
{\rm Mass}_{(\bmu,\balpha,{\bf b})} \sim \left|\bmu\cdot{\bf a}+
\balpha\cdot{\bf a}^D +b_A m_A \right|
\label{BPSmass}
\ee
are never simultaneously real, except for singular or degenerate cases, when some of these period(s)
vanish, giving rise to the extra massless states in the spectrum. These massless
states lead to possible decays, causing change of the quantum numbers for the light states,
and therefore
in different domains of the moduli space the condensates acquire different charges.
In order to determine these charges one has to consider the Seiberg-Witten theory for supersymmetric QCD in the vicinity of singular curves, corresponding to \none vacua.

\section{Seiberg-Witten theory for supersymmetric QCD
\label{ss:swfund}}

Generic curve for \ntwo supersymmetric QCD with $N_c$ colors and $N_f$ flavors can be written
in the form \cite{HO,GMMM}
\be
\label{cuy}
y^2=P(x)^2-4Q(x)
\ee
where
\be
\label{PQ}
P(x)=\prod_{i=1}^{N_c}(x-\phi_i), \ \ \
\sum_{i=1}^{N_c}\phi_i=-\Lambda\delta_{N_f,2N_c-1}
\\
Q(x)=\Lambda^{2N_c-N_f}\prod_{A=1}^{N_f}(x+m_A)
\ee
are two polynomials of powers $N_c$ and $N_f$ respectively.
In semiclassical regime the roots $\{ \phi_i \}$, $i=1,\ldots,N_c$ of the first one
coincide
with the eigenvalues of matrix $\Phi$ of the condensate of the
complex scalar from the vector multiplet of
\ntwo supersymmetric Yang-Mills theory, but being computed exactly
they (or, better, their symmetric functions) are got corrected in
(dependent upon $\Lambda$ and $m_A$ way) due to the instanton
effects.

The hyperelliptic representation \rf{cuy} can be also re-written in the form
\be
\label{cuw}
w+{Q(x)\over w}=P(x)
\ee
or\footnote{Sometimes the asymmetric representation with the factorized $Q(x)=Q_+(x)Q_-(x)$
and $w={\sf w}Q_+(x)$, so that equation \rf{cuw} turns into ${\sf w}^2Q_+(x)-P(x){\sf w}+Q_-(x)=0$ is more adequate from the point of view of brane-construction \cite{Gai} and
possible relations to the matrix models \cite{EgM} and quantum integrable systems.}
\be
\label{cuW}
W+{1\over W}={P(x)\over\sqrt{Q(x)}}
\ee
with $y=w-{Q(x)\over w} = \left(W-{1\over W}\right)\sqrt{Q(x)}$.

The curves \rf{cuy}, \rf{cuw} or \rf{cuW} are endowed with a generating differential
\be
\label{dS}
dS \sim x{dw\over w} = x{dW\over W}+ \frac12 x{dQ\over Q} = {x
dP\over y} - x {P\over 2y}{dQ\over Q}+ \frac12 x{dQ\over Q}
\ee
whose periods \rf{SWper} enter the mass formula \rf{BPSmass}, as well as its
residues
\be
\label{resdS}
\res_{P_A^\pm}dS =  m_A \cdot \left.{P\over 2y}\right|_{x=-m_A}-{m_A\over 2} =
%\pm \half
-m_A
\ee
at the points $P_A%^\pm
$ with $x(P_A^\pm)=-m_A$ at one of the sheets of \rf{cuy}. The residues \rf{resdS}
must disappear in the
limit of vanishing masses \cite{SW2}, and for $N_f=2N_c-1$ this requirement
causes nonvanishing term of the order of $\Lambda$ in the r.h.s. of the second
equality in \rf{PQ}.

The variation of
\rf{dS} at constant $W$ gives rise to
\be
\label{vardS}
\delta dS \sim {dx\over y}\left(\delta P(x) -
\half P{\delta Q(x)\over Q(x)}\right)
\ee
In the case of $SU(3)$ gauge group it is convenient to introduce explicitly
\be
P(x)=(x-\phi_1)(x-\phi_2)(x-\phi_3) = x^3-ux-v
\ee
with, for $N_f<5$ and $\phi_3=-\phi_1-\phi_2$, so that
\be
\label{uvphi}
u=\phi_1^2+\phi_2^2+\phi_1\phi_2
\\
v=-\phi_1\phi_2(\phi_1+\phi_2)
\ee
Now we turn to particular cases of these formulas in the vicinity of singular curves.

\section{Period integrals on degenerate curves
\label{ss:dgcu}}

\subsection{$N_c=2$ and $N_f=2$ case
\label{ss:n2per}}

Starting originally with the $SU(3)$ gauge theory around vacuum with two condensed flavors
($r=2$ in terms of \cite{MY1}) at weak coupling, and moving it towards the strongly coupled domain, one gets
the picture, effectively described by $SU(2)$ gauge theory with $N_f=2$ light
flavors, when $r=2$ vacuum collides with the vacuum, containing also massless
monopole \cite{MY2}.
In this warmup example with $N_c=2$ and $N_f=2$, taken for simplicity with the coinciding
masses $m_1=m_2=m$, one gets for the curve \rf{cuy}
\be
\label{n2n2}
y^2 = \left( x^2-u\right)^2 - 4\Lambda^2(x+m_1)(x+m_2) =
\left( x^2-u\right)^2 - 4\Lambda^2(x+m)^2
\ee
while the generating differential \rf{dS} turns for the pairwise
coinciding masses into
\be
dS \sim {x dP\over y} - x {P\over 2y}{dQ\over Q}+ \frac12 x{dQ\over
Q} = {x dP\over y} - x {P\over y}{dq\over q}+ x{dq\over q}
\\
q(x) = \prod_{B=1}^{N_f/2}(x+m_B)
\label{dSm2}
\ee
where we have chosen $m_{B+N_f/2}=m_B$, $B=1,\ldots,N_f/2$ for even
number of flavors $N_f$. The curve \rf{n2n2} just corresponds to
$P(x)=x^2-u$ and $q(x)=x+m$.

When exactly at quark vacuum - i.e. at the point $u=u_Q=m^2$, the curve \rf{n2n2}
degenerates further to
\be
y^2 =
(x+m)^2\left((x-m)^2-4\Lambda^2\right)
\equiv (x+m)^2Y^2
\\
Y^2= (x-m)^2-4\Lambda^2
\label{n2n2q}
\ee
and the Seiberg-Witten differential \rf{dS} turns into
\be
dS = {xdx\over Y}+{xdx\over x+m}
\label{dSm2q}
\ee
where the first term in the r.h.s. coincides with the generating differential
for the formal
pure \ntwo SUSY $U(1)$ gauge theory, with the only VEV given by $m$
\cite{LMN,MN}.

Due to \rf{resdS} and to the fact, that on degenerate curve the
position of the mass pole at $x=-m$ (for both flavors) coincides
with the degenerate cut, the differential \rf{dSm2q} is normalized
by
\be
{1\over 2\pi i}\oint_{x=-m} dS_+ = {1\over 2\pi i}\oint_{A^+} dS_+ =
a = -m
\\
{1\over 2\pi i}\oint_{x=-m} dS_- = \res_{x=-m}dS_- - {1\over 2\pi
i}\oint_{A^-} dS_- = -2m+a = -m
\label{resdSm2q}
\ee
obviously true for \rf{dSm2q}. Since the curve
\rf{n2n2q} is rational, the differential
\rf{dSm2q} can be easily integrated, giving rise to
\be
S = Y + m\log(x-m+Y)+ x -m\log(x+m)
\label{Sm2q}
\ee
In order to compute the desired $B$-period (the monopole mass), one
has to take the difference $\left.S_+\right|_{x=-m} -
\left.S_-\right|_{x=-m}$ of the values of \rf{Sm2q} on two different
sheets of the Riemann surface \rf{n2n2q}.

This is not possible to do by direct substitution of $x=-m$ into
\rf{Sm2q} due to the logarithmic singularity, i.e. the curve
\rf{n2n2q} is ``too degenerate''\footnote{For example, already the period matrix of \rf{n2n2q}
is not a well-defined object. One can verify, that at the Argyres-Douglas point, when $m=\Lambda$,
we get $\tau_\ast = i$, as follows from the self-duality requirement \cite{BF,AD}, since
the curve \rf{n2n2q} degenerates to $y^2\sim (x+m)^3$, giving rise to the desired limiting value,
see e.g. \cite{BE}.}.
Let us then slightly regularize it, and denote
the distance between the position of the pole and the nearest end of
the shrinked cut by $\epsilon^\pm$, dependently on the sheet
$Y=Y_\pm$ of \rf{n2n2q}. The values of these
$\epsilon^\pm=\epsilon^\pm(m,\Lambda)$ can be determined as follows
(see e.g. \cite{MN}): the differential
\be
d\phi = {dw\over w}\ \stackreb{\rf{n2n2q}}{=}\ {dx\over Y}+{dx\over
x+m}
\label{dphi}
\ee
should have constant periods \cite{KriW}, moreover, its $B$-periods
on \rf{n2n2q} can be just chosen vanishing. Integrating \rf{dphi} up
to
\be
\phi = \log (x-m+Y) + \log (x+m)
\label{phi}
\ee
and putting $\left.\phi_+\right|_{x=-m} -
\left.\phi_-\right|_{x=-m} \equiv \left.\phi_+\right|_{x=-m+\epsilon^+} -
\left.\phi_-\right|_{x=-m+\epsilon^-}=0$, one gets (at
$\epsilon^\pm\to 0$)
\be
\log{\epsilon^+\over\epsilon^-} =
\log {m+\sqrt{m^2-\Lambda^2}\over m-\sqrt{m^2-\Lambda^2}}
\label{eml}
\ee
Therefore
\be
\left.S_+\right|_{x=-m} - \left.S_-\right|_{x=-m} =
-m\log{\epsilon^+\over\epsilon^-} + 2\left.Y\right|_{x=-m} +
m\log{-2m + \left.Y\right|_{x=-m}\over -2m - \left.Y\right|_{x=-m}}
=
\\
= 4\sqrt{m^2-\Lambda^2}+  2m\log {m-\sqrt{m^2-\Lambda^2}\over
m+\sqrt{m^2-\Lambda^2}}
\label{SD}
\ee
Hence, evaluating $B$-period on degenerated curve
\rf{n2n2q} gives rise to explicit formula (cf. the result with \cite{Dorey}, where
similar structures have been obtained by indirect methods from two-dimensional approach)
\be
a^D = {1\over 2\pi i}\left(\left.S_+\right|_{x=-m} -
\left.S_-\right|_{x=-m}\right) = -{i\over
\pi}\left(2\sqrt{m^2-\Lambda^2}+m
\log{m-\sqrt{m^2-\Lambda^2}\over m+\sqrt{m^2-\Lambda^2}}\right)
\label{aDm}
\ee
showing in particular, that $\left.\Im(a^D)\right|_{m=\pm \Lambda}=0$.

However, to analyze the difference between massless monopoles and dyons
one should also consider carefully the real part of \rf{aDm}, taking
into account the logarithmic cut in $m$-plane. One can fix this real part to vanish at
$m=\Lambda$, then
\be
\left.\Re (a^D)\right|_{m=\Lambda}=0
\\
\left.\Re (a^D)\right|_{m=-\Lambda} = 2m = -2a
\label{reaDm}
\ee
It means, that when quark singularity $u_Q$ collides with $u_M$ we
get massless monopole with $|a^D|=0$, while when $u_Q$ collides
with $u_D$, one gets the vanishing mass of the $(1,1)=[2,1]$ dyon,
$|a^D+2a|=0$. Here, following \cite{SW2,MY2}, we specially point out, that
the ``homological'' electric charge $[2,1]$ of this dyon is different from conventional
physical one $(1,1)$, while the magnetic charges coincide in both normalizations.
The electric charge of this dyon coincides with the
charge of W-boson and {\em doubles} the charge of a quark: the last one is
$(\half,0)$ in physical normalization, but corresponds to single $A$-cycle (i.e.
equals to $[1,0]$) in homological one, therefore the electric charge of
this dyon (and of the W-boson) corresponds to {\em two} $A$-cycles.

Asymptotically for \rf{aDm} one has
\be
a^D\ \stackreb{m\to\infty}{\simeq}\ {i\over\pi} 2m\left(
\log{2m\over\Lambda}-1\right) + \ldots
\label{aDmas}
\ee
For the mass-derivatives of \rf{aDm} one gets
\be
{\d a^D\over \d m} = - {i\over\pi}
\log{m-\sqrt{m^2-\Lambda^2}\over m+\sqrt{m^2-\Lambda^2}}
\label{tn2n2}
\ee
and
\be
{\d^2 a^D\over \d m^2}={2i\over\pi\sqrt{m^2-\Lambda^2}}
\label{dta}
\ee
When the second derivative
diverges, one can write the fractional-power expansions:
\be
a^D\ \stackreb{m\to\Lambda}{\simeq}\
{4i\over 3\pi}\sqrt{2\over\Lambda}\cdot (m-\Lambda)^{3/2} + \ldots
\label{aDL}
\ee
at $m=\Lambda$ and, similarly
\be
a^D\ \stackreb{m\to-\Lambda}{\simeq}\ 2m
+{4\over 3\pi}\sqrt{2\over\Lambda}\cdot (m+\Lambda)^{3/2} + \ldots
\label{aDmL}
\ee
at $m=-\Lambda$. Expressions \rf{aDL} and \rf{aDmL} can be easily analyzed for
the presence of nontrivial monodromies in the mass plane.

In particular, one finds from \rf{aDL}, that being pure imaginary at $t=m-\Lambda>0$, when
continued to negative value $t=m-\Lambda<0$ the value of the period $a^D$ is either
positive or negative dependently on the chosen way in the mass plane, or the sign of the angle
variable in $t=\varepsilon e^{\pm i\varphi}$ at $t\approx 0$. This formula says, that
\be
a^D \approx {4i\over 3\pi}\sqrt{2\over\Lambda}\ t^{3/2} = \pm
{4\over 3\pi}\sqrt{2\over\Lambda}\ \varepsilon^{3/2}
\label{aDt}
\ee
Suppose one now takes $u=m^2+\delta$ for $m\gg |\delta|>0$ in the vicinity of
quark vacuum. Then
$a=-\sqrt{u} \simeq -m - {\delta\over 2m}$, or $a+m \simeq - {\delta\over 2m}$.
When $m$
approaches $\Lambda$ at $t\to 0$ the mass of the light quark is
\be
|a+m|\simeq \left|{\delta\over 2m}\right| \simeq \left|{\delta\over
2\Lambda}\right| \gtrsim 0
\label{pmq}
\ee
Going to
negative $t$ formula \rf{aDt} says that, dependently on the chosen way in $m$-plane,
after crossing the critical line the mass of one of the $(\half,\pm 1)=[1,\pm 1]$ dyons
becomes less than the mass of the quark. These conditions can be formalized as
\be
\label{signs}
{\rm sign}(\Im\ t) = {\rm sign}(\delta):\ \ \ \ \ \ \
|a+m + a^D| < |a+m|
\\
{\rm sign}(\Im\ t) = -{\rm sign}(\delta):\ \ \ \ \
|a+m - a^D| < |a+m|
\ee
If it happens, say, for the positive magnetic charge (i.e. the positive sign in front of
$a^D$) in the inequality $|a+m + a^D| < |a+m|$,
the quark can
emit the massless anti-monopole and turn into the $(\half,1)=[1,1]$
dyon with the charge conservation law $[1,0]+[0,1]=[1,1]$, in the opposite
case the sign of the monopole charge has to be changed, i.e. generally one gets
\be
\label{proc}
[1,0]+[0,q]=[1,q], \ \ \ \ q=\pm 1
\ee
and the particular choice of the sign $q$, as we see below, is not observable.
Similarly, at another critical line, at $m=-\Lambda$, one gets the
process with the conservation law $[1,q]=[2,q]+[-1,0]$, with $q=\pm 1$ again,
or the dyon $[1,q]$ decouples into the
$[2,q]$-dyon, massless at $m=-\Lambda$ due to \rf{reaDm}, and the
quark $[-1,0]$, which can be treated as ``dual to''
$[1,0]$-quark after exchange the sign of the mass $m\leftrightarrow
-m$, see \cite{MY2}.

The analysis of this section can be supplemented by direct consideration
of the process of permutation of the branch points of the Seiberg-Witten curve
\rf{cuy} in the vicinity of the singularities \cite{MY2,SYdual}, which leads,
basically, to the same conclusion.

\subsection{$N_c=3$ and $N_f=4,5$ theories}

Let us now turn directly to the $SU(3)$ theory, with
$N_f=4$ with the pairwise coinciding masses $m_1=m_3$ and $m_2=m_4$,
when the curve \rf{cuy} becomes
\be
\label{su3nf4}
y^2=
(x-\phi_1)^2(x-\phi_2)^2(x+\phi_1+\phi_2)^2-4\Lambda^2(x+m_1)^2(x+m_2)^2
\ee
Putting exactly $\phi_i=-m_A\delta_{i,A}$ for $i,A=1,2$ for the
chosen quark vacuum, the curve \rf{su3nf4} degenerates into
\be
\label{nf4deg}
y^2 = (x+m_1)^2(x+m_2)^2\left((x-m_1-m_2)^2-4\Lambda^2\right) =
\\
= (x+m_1)^2(x+m_2)^2\left((x-M)^2-4\Lambda^2\right)\equiv
(x+m_1)^2(x+m_2)^2Y^2
\\
Y^2 = (x-M)^2-4\Lambda^2
\\
M=m_1+m_2
\ee
The Seiberg-Witten differential \rf{dS} turns in this case into
\be
dS
%= {xdp\over Y}+\frac12 {dQ\over Q}
= {xdx\over Y}+{xdx\over x+m_1}+{xdx\over x+m_2}
\\
p = x-M,\ \ \ Y^2 = p^2-4\Lambda^2
\label{dSu1}
\ee
The reasoning for its normalization
\be
{1\over 2\pi i}\oint_{x=-m_k} dS_+ = {1\over 2\pi i}\oint_{A_k^+}
dS_+ = a_k = -m_k
\\
{1\over 2\pi i}\oint_{x=-m_k} dS_- = \res_{x=-m_k}dS_- - {1\over
2\pi i}\oint_{A_k^-} dS_- = -2m_k+a_k = -m_k
\\
k=1,2
\label{resdSsu3}
\ee
just literally repeats that of \rf{resdSm2q} for each mass $m_{1,2}$
in \rf{nf4deg}, \rf{dSu1}. Clearly, the residues \rf{resdSsu3}
corresponds to exact vanishing of the effective masses of the quarks
$a_K+m_K=0$ ($K=1,2$) in the chosen quark vacuum with two condensed flavors.

Let us turn to the dual to \rf{resdSsu3} $B$-periods, corresponding via
$\bmu_I\cdot\balpha_J=\delta_{IJ}$ ($I,J=1,2$) to the monopole
masses. Again, they are given by the differences
$\left.S_+\right|_{x=-m_k} -
\left.S_-\right|_{x=-m_k}$ ($k=1,2$) of the values of the Abelian
integral of \rf{dSu1}
\be
S = Y + M\log(x-M+Y) + 2x -m_1\log(x+m_1) -m_2\log(x+m_2)
\\
M=m_1+m_2
\label{Ssu3}
\ee
Since, this is again singular at $x+m_k=0$ one needs to introduce the
regulators $\epsilon^\pm_k$, $k=1,2$, on each sheet of \rf{nf4deg}.
As in \rf{eml} they are determined by
\be
\left.\phi_+\right|_{x=-m_k} - \left.\phi_-\right|_{x=-m_k}=
\log{\epsilon_k^+\over\epsilon_k^-} +
\log {M+m_k-\sqrt{(M+m_k)^2-4\Lambda^2}\over M+m_k+\sqrt{(M+m_k)^2-
4\Lambda^2}} = 0
\\
k=1,2
\label{ek}
\ee
where
\be
\phi = \int \left({dx\over Y}+{dx\over x+m_1}+{dx\over
x+m_2}\right)=
\\
 = \log(x-M+Y) + \log(x+m_1) + \log(x+m_2)
\label{phisu3}
\ee
Thus
\be
\left.S_+\right|_{x=-m_k} - \left.S_-\right|_{x=-m_k}
= -m_k\log{\epsilon_k^+\over\epsilon_k^-} + 2\left.Y\right|_{x=-m_k}
+ M\log{-m_k-M + \left.Y\right|_{x=-m_k}\over -m_k-M -
\left.Y\right|_{x=-m_k}} =
\\
= 2\sqrt{(m_k+M)^2-4\Lambda^2}+  (m_k+M)\log
{m_k+M-\sqrt{(m_k+M)^2-4\Lambda^2}\over
m_k+M+\sqrt{(m_k+M)^2-4\Lambda^2}}
\\
k=1,2
\label{SDk}
\ee
Hence, (cf. with \rf{aDm}),
\be
\label{B1dg}
a_1^D = {1\over 2\pi i}\oint_{B_1}dS =
\\
= -{i\over \pi}\left(\sqrt{(2m_1+m_2)^2-4\Lambda^2}
+\left(m_1+{m_2\over
2}\right)\log{2m_1+m_2-\sqrt{(2m_1+m_2)^2-4\Lambda^2}\over
2m_1+m_2+\sqrt{(2m_1+m_2)^2-4\Lambda^2}}\right)
\ee
and
\be
\label{B2dg}
a_2^D = {1\over 2\pi i}\oint_{B_2}dS =
\\
= -{i\over \pi}\left(\sqrt{(m_1+2m_2)^2-4\Lambda^2} +\left({m_1\over
2}+m_2\right)\log{m_1+2m_2-\sqrt{(m_1+2m_2)^2-4\Lambda^2}\over
m_1+2m_2+\sqrt{(m_1+2m_2)^2-4\Lambda^2}}\right)
\ee
which obviously obey the desired properties, similarly to $N_c=2$,
$N_f=2$ theory, which effectively describes dynamics in the subgroups of the
$SU(3)$ gauge group, corresponding to the roots $\balpha_{1,2}$.

In the $N_f=5$ theory, the derivation is quite similar, though the
formulas are slightly more complicated. The degenerate curve is again
\be
\label{nf5dg}
y^2= (x+m_1)^2(x+m_2)^2Y^2
\ee
where now
\be
\label{u1n1}
Y^2 = p^2 - 4\Lambda(x+m_5)
\\
p = x-M+\Lambda = x-m_1-m_2+\Lambda
\ee
and it is endowed with the Seiberg-Witten differential \rf{dS} to be now
\be
dS = {xdx\over Y} - {xp\over 2Y}{dx\over x+m_5} + \frac12 {xdx\over
x+m_5} + {xdx\over x+m_1}+ {xdx\over x+m_2}
\label{dgswnf5}
\ee
Computation of the $B$-periods of the differential
\rf{dgswnf5}, similar to the $N_f=4$ case, leads to the result
\be
\label{Bkdg5}
a_k^D = {1\over 2\pi i}\oint_{B_k}dS =
 -{i\over \pi}\left(Y_k +\left(M+{m_k+m_5\over
2}\right)\log{M+m_k+\Lambda-Y_k\over M+m_k+\Lambda+Y_k} + \right.
\\
\left. +
{m_k-m_5\over 2}\log{a+b_+m_k-b_-Y_k\over a+b_+m_k+b_-Y_k}
\right),\ \ \ k=1,2
\\
a=(M-\Lambda)^2+m_5(M-3\Lambda),\ \ \ b_\pm=M+m_5\pm\Lambda
\ee
with two parabolic curves $Y_{1,2}$ in the $(m_1,m_2)$-mass plane, being defined by
\be
Y_1^2=(2m_1+m_2)^2 - \Lambda(2m_5+m_2)+ \Lambda^2=0
\label{par1}
\ee
and
\be
Y_2^2=(m_1+2m_2)^2 - \Lambda(2m_5+m_1)+ \Lambda^2=0
\label{par2}
\ee
These parabolic curves, appearing naturally in the process of the period computation,
play the role of the borders of the deformed strongly coupled domain in the $N_f=5$ case,
compare to the $N_f=4$ case, where these borders are just straight lines $m_1+2m_2=\pm 2\Lambda$
and $2m_1+m_2=\pm 2\Lambda$.

It is easy to check, that \rf{Bkdg5} obeys all desired properties.
Note also, that this derivation has nothing in common with that of \cite{Dorey,SYdual},
where the indirect methods, based on parallels between four-dimensional and two-dimensional
approaches have been exploited.

Formulas \rf{B1dg}, \rf{B2dg} and \rf{Bkdg5} immediately lead to simple physical
conclusions \cite{MY2}. Vanishing of $a^D_k=\balpha_k\cdot{\bf a}^D$, $k=1,2$, corresponds to vanishing of masses of the monopoles with the charges
\be
\label{monchar}
M_k = \sqrt{2}\left({\bf n}_e\oplus{\bf n}_m\right)^M_k = {\bf 0}\oplus\balpha_k,
\ \ \ \ k=1,2
\ee
in terms of the roots of the $SU(3)$ gauge group, see fig.~\ref{fi:su3}.
This happens for the period \rf{B1dg} at $2m_1+m_2=2\Lambda$, and for \rf{B2dg}
at $m_1+2m_2=2\Lambda$.

Clearly, the imaginary part $\Im(a_1^D) =0$ vanishes also at
$2m_1+m_2=-2\Lambda$, and similarly $\Im(a_2^D) =0$ if
$m_1+2m_2=-2\Lambda$. However, the real parts of expression
\rf{B1dg} at $2m_1+m_2=-2\Lambda$ equals to $2m_1+m_2 = -
\balpha_1\cdot{\bf a}$, or to the mass of the W-boson with the charge $\balpha_1$.
Hence, at $2m_1+m_2=-2\Lambda$ one gets the massless dyon with the
charge
\be
D_1 = \sqrt{2}\left({\bf n}_e\oplus{\bf n}_m\right)^D_1 = \balpha_1\oplus\balpha_1
\label{a1dyc}
\ee
and similarly, the massless dyon with the charge
\be
D_2 = \sqrt{2}\left({\bf n}_e\oplus{\bf n}_m\right)^D_2 = \balpha_2\oplus\balpha_2
\label{a2dyc}
\ee
at $m_1+2m_2=-2\Lambda$. Formulas
\rf{Bkdg5} show that at $Y_k=0$, $k=1,2$ (i.e.
for each parabola \rf{par1}, \rf{par2} in the mass plane) the
imaginary part of the corresponding period $\Im(a^D_k)$ vanishes,
while the real part $\Re(a^D_k)$ jumps when passing from the
positive to negative branch of the corresponding $k$-th parabola.

%%%%%%%%%%%%%%%%%%%%%%%%%%%%%%%%%%%%%%%%%%%%%%%%%%%%%%%%%%%%%%%%%%%%%%%%%%%%%%%%%%%%%%%%%%%%%%%%
\begin{figure}[tp]
\epsfysize=5cm
\centerline{\epsfbox{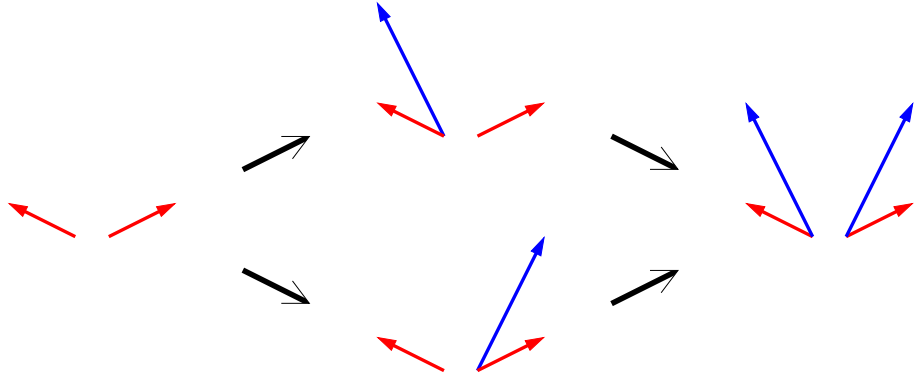}}
\caption{Change of the quantum numbers due to monodromies:
the doublet of quarks (weak coupling) turns into the doublet of dyons (at strong
coupling), dependently on the way chosen in the mass-plane.}
\label{fi:qua_dyo}
\end{figure}
%%%%%%%%%%%%%%%%%%%%%%%%%%%%%%%%%%%%%%%%%%%%%%%%%%%%%%%%%%%%%%%%%%%%%%%%%%%%%%%%%%%%%%%%%%%%%%%%%

The analysis of the computed periods \rf{B1dg}, \rf{B2dg} and \rf{Bkdg5} leads to
the following picture of changing of the quantum numbers due to monodromies (see
fig.~\ref{fi:qua_dyo}): starting with the condensate of two quarks at weak coupling
one ends up in strongly-coupled phase with the condensate of two light dyons. Formula
\rf{proc} of the previous section describes the projection of fig.~\ref{fi:qua_dyo}
to the horizontal line: in terms of the weight and root vectors of fig.~\ref{fi:su3}
one finds that each quark with the weight-like electric charge pickups up the
root-like magnetic charge and turns into a dyon of {\em different} nature from \rf{a1dyc}
and \rf{a2dyc}: in the strongly-coupled phase the light dyons $\Psi_{1,2}$ with the
charges\footnote{
Let us point out here, that by calligraphic ${\cal D}$-letters we denote the dyons
with the weight-like electric charges, in contrast to the dyon-relatives of the monopoles
with the charges \rf{a1dyc} and \rf{a2dyc}, given entirely in terms of the root vectors.}
\be
{\cal D}_K  = \bmu_K\oplus\balpha_K =
\sqrt{2}\left({\bf n}_e\oplus{\bf n}_m\right)^{\cal D}_K \equiv \sqrt{2}{\bf n}^K,
\ \ \ \ \ \ K=1,2
\label{dyon12}
\ee
condense instead quarks \cite{MY2}.

\section{Confinement}

Since in the weak coupling regime of the original theory at large $m$ in
$r=2$ vacuum the quarks are in the Higgs phase, they confine the monopoles.
Two of three $SU(3)$ elementary monopoles with the magnetic charges
$\balpha_{1,2}$  (see
\rf{monchar}) are attached to the ends of the elementary strings, while the
third one with the charge $\balpha_{12}=\balpha_1-\balpha_2$
becomes a string junction of two elementary strings
\cite{MY1}. At strong coupling one can repeat these reasoning for the dual theory
of light dyons \rf{dyon12}, where the fundamental strings can be constructed
from the effective Lagranian:
${\cal L} \sim \sum_{K=1,2}\left|\nabla \Psi_K\right|^2 + \ldots$, where the
minimal interaction
\be
\label{minint}
\nabla \Psi_K = \left(\d - i{\cal D}_K\cdot({\bf A}\oplus{\bf A}^D)\right)\Psi_K = \left(\d - i(\bmu_K\cdot{\bf A}+\balpha_K\cdot{\bf A}^D)\right)\Psi_K
\ee
is determined by the dyon charges \rf{dyon12}.

The elementary charges $S_{1,2}$ of the $\mathbb{Z}_2$-strings follow from the dyon charges
\rf{dyon12} via behavior of the gauge
potentials at spatial infinity,
\be
\label{StDy}
S_I:\ \ \ \ \ {\cal D}_K\cdot({\bf A}\oplus{\bf A}^D) =
\bmu_I\cdot{\bf A}+\balpha_I\cdot{\bf A}^D \sim \delta_{IK}d\theta,\ \ \ I,K=1,2
\ee
where $\theta$ is angle in the plane, transverse to the direction of string.
This implies for the Cartesian projections $A_3 \sim \balpha_{12}\cdot{\bf A} \sim (\bmu_1-\bmu_2)\cdot{\bf A}$
and $A_8 \sim (\balpha_1+\balpha_2)\cdot{\bf A}\sim -\bmu_{12}\cdot{\bf A}$ (the horizontal
and vertical directions at fig.~\ref{fi:su3}, while the indices $\{3,8\}$ come from the diagonal Gell-Mann matrices)
\be
 A_3 + A_3^{D}  \sim  d\theta,
\ \ \ \ \
   \frac{A_8}{\sqrt{3}}
+ \sqrt{3} A_8^{D} \sim d\theta
\label{gw}
\ee
The combinations orthogonal to (\ref{gw}) should vanish at infinity
\be
\label{gwo}
A_3 - A_3^{D}\sim 0,\ \ \ \ \
A_8^{D} -3A_{8} \sim 0
\ee
As a result \cite{MY2} one gets for \rf{StDy} in terms of the fluxes\footnote{
This definition ensures that the string has the same charge as a
trial dyon which can be attached to the string endpoint (not
necessarily being present in the spectrum of the theory).}
\be
\oint dx\cdot({\bf A}^{D}\oplus {\bf A}) =\oint dx\cdot(A_3^{D},A_3;A_8^{D},A_8)=
\\
= 4\pi\,(-n_3^e,n_3^m;-n_8^e,n_8^m) = 4\pi (-{\bf n}^{e}\oplus {\bf n}^m)
\label{defstrch}
\ee
that the charge of the $S_1$-string is
\be
{\bf n}_{S_1}=
\left(\,-\frac14,\,\frac14;\,-\frac{3\sqrt{3}}{20},\,\frac{\sqrt{3}}{20}\right)
\equiv {S_1\over\sqrt{2}}
\label{S1}
\ee
while the charge of the $S_2$-string, arising due to winding at spatial
infinity of the phase of the second dyon, equals to
\be
{\bf n}_{S_2}=
\left(\,\frac14,\,-\frac14;\,-\frac{3\sqrt{3}}{20},\,\frac{\sqrt{3}}{20}\right)
\equiv {S_2\over\sqrt{2}}
\label{S2}
\ee
Now one can easily check that each of three $SU(3)$ monopoles can be indeed
confined by these two strings \cite{MY2}. For the monopoles with the
charges ${\bf 0}\oplus\balpha_{1,2}$ or
$(\,0,\pm\frac12;\,0,\frac{\sqrt{3}}{2})$ one has the following decompositions
\be
{1\over\sqrt{2}}({\bf 0}\oplus\balpha_1) =
(\,0,\frac12;\,0,\frac{\sqrt{3}}{2})= {\bf n}_{S_1} +\frac7{10} {\bf
n}^1+\frac2{10}{\bf n}^2,
\\
{1\over\sqrt{2}}({\bf 0}\oplus\balpha_2) =
(\,0,-\frac12;\,0,\frac{\sqrt{3}}{2})= {\bf n}_{S_2} +\frac2{10}
{\bf n}^1+\frac7{10}{\bf n}^2
\label{mondec}
\ee
where ${\bf n}^K={{\cal D}_K\over\sqrt{2}}$, $K=1,2$ are the normalized charges of the
dyons \rf{dyon12}. Formula \rf{mondec}
shows, that only a part of the monopole-anti-monopole flux is confined to the
interior of the string world-sheet, while the remaining part is just screened by the dyon condensate.

For the third $SU(3)$ $\balpha_{12}$-monopole the formulas are more elegant:
one gets from
(\ref{mondec}), that
\be
{1\over\sqrt{2}}({\bf 0}\oplus\balpha_{12}) = (0,1;\,0,0)= {\bf
n}_{S_1} - {\bf n}_{S_2} +\frac12\left( {\bf n}^1-{\bf n}^2\right)
\label{conf12}
\ee
and we find that it is also confined, being a junction of two
elementary strings $S_1$ and $S_2$.
Pictorially, dynamics in the non-Abelian direction $\balpha_{12}$ is determined by
``difference dyon''
${\cal D}_1-{\cal D}_2\ =\ $ \epsfysize=0.3cm\epsfbox{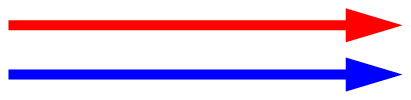}
and ``difference string'' $S_1-S_2=$ \epsfysize=0.3cm\epsfbox{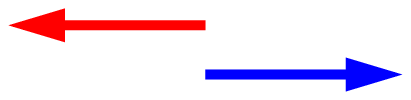}. Formula
\rf{conf12} claims, that due to the total
screening of the electric charge of the ``difference string'' by the condensate of the
``difference dyon''
$S_1-S_2 + \half({\cal D}_1-{\cal D}_2)\ =\ \epsfysize=0.16cm\epsfbox{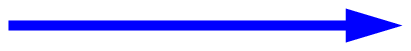}$,
the confinement concerns only the magnetic charges of the ($SU(2)$)-monopoles.

Let us now come back to the problem of ambiguity.
Note, that when passing to the strong coupling domain, the sign
of magnetic charge, being absorbed by light quark turning into a
light dyon is not observable, since it depends on the choice of
particular trajectory in the space of masses (see discussion in
sect.~\ref{ss:n2per}). This means, that instead of the theory of
light dyons with the charges \rf{dyon12} one could also
consider the theory, where dyons ${\tilde{\cal D}}_{1,2}$ have the charges
\be
\sqrt{2}\tilde{\bf n}^1=\bmu_1\oplus(-\balpha_1),\ \ \ \ \
\sqrt{2}\tilde{\bf n}^2=\bmu_2\oplus(-\balpha_2)
\label{tdc}
\ee
Such theory is in fact equivalent to have been just discussed above:
basically only the signs of the components of the dual gauge
fields $A^D$ have to be changed for the opposite. This results, instead of
\rf{S1} and \rf{S2}, in the elementary string charges
\be
\tilde{\bf n}_{S_1}=
\left(\frac14,\,\frac14;\frac{3\sqrt{3}}{20},\,\frac{\sqrt{3}}{20}\right)
\label{tS1}
\ee
and
\be
\tilde{\bf n}_{S_2}=
\left(-\frac14,\,-\frac14;\frac{3\sqrt{3}}{20},\,\frac{\sqrt{3}}{20}\right)
\label{tS2}
\ee
i.e. they will come with the opposite electric components. However, one can easily
check, that instead of \rf{mondec} one can now write the following decomposition
\be
{1\over\sqrt{2}}({\bf 0}\oplus\balpha_1) =
(\,0,\frac12;\,0,\frac{\sqrt{3}}{2})= \tilde{\bf n}_{S_1}
-\frac7{10} \tilde{\bf n}^1-\frac2{10}\tilde{\bf n}^2,
\\
{1\over\sqrt{2}}({\bf 0}\oplus\balpha_2) =
(\,0,-\frac12;\,0,\frac{\sqrt{3}}{2})= \tilde{\bf n}_{S_2}
-\frac2{10} \tilde{\bf n}^1-\frac7{10}\tilde{\bf n}^2
\label{conftS}
\ee
and we find that the main conclusion of \cite{MY2}
remains intact under the change
of the sign of magnetic components of the dyon charges: the
monopoles are confined by the strings in effective theory, with the
part of their fluxes being screened by the dyon condensate. It means
in particular, that dependently on choosing a particular way in the mass
plane when
going from weak to strongly coupled domain of the original theory
one gets different charges of the condensed dyons and string solutions
in effective theory. However, this ambiguity does not influence to the
observable physical conclusion: in any case {\em the same} magnetic monopoles are
confined.

\section{Conclusion}

We have discussed here some details of the Seiberg-Witten theory for degenerate
curves in the vicinity of \none vacua for the supersymmetric QCD. The computation
of periods on these singular curves allow to understand, how the quantum numbers of
light states change, when passing from the weak to strong-coupled domains by choosing
some trajectory in the mass moduli space. This analysis helps to get an exact picture of
confinement in supersymmetric QCD at strong coupling.

It has been shown above, that the computation of these periods can be performed directly,
using the technique of integrable systems. In this framework it simply means, that the
exact solution to supersymmetric gauge theory is characterized by a curve with two meromorphic
differentials with the fixed periods. Fixing these periods is enough to provide regularization
of the singular curve, corresponding to a \none vacuum, and therefore the masses of the light states can be exactly computed. This is one more nontrivial application of the classical integrable systems in quantum theory.

\bigskip\bigskip\noindent
This work was supported by the Russian Federal Nuclear Energy
Agency, by Grant of Support for Scientific
Schools LSS-1615.2008.2, by the RFBR grants 08-01-00667,
09-02-90493-Ukr, 09-02-93105-CNRSL, 09-01-92440-CE, by Russian Ministry of Education and Science under the contract 02.740.11.0608, and by Kyoto
University. I would like to thank
the Yukawa Institute for Theoretical Physics, where the most essential part
of the work has been done, for the warm hospitality.

%\newpage


\begin{thebibliography}{7799}
%\bigskip\bigskip

\bibitem{MY1}
 A.~Marshakov and A.~Yung,
  %``Non-Abelian confinement via Abelian flux tubes in softly broken N = 2  SUSY
  %QCD,''
  Nucl. Phys. {\bf B647} (2002) 3,
  [arXiv:hep-th/0202172].

\bibitem{MY2}
A.~Marshakov and A.~Yung,
  %``Strong versus Weak Coupling Confinement in N=2 Supersymmetric QCD,''
  Nucl. Phys. {\bf B 831} (2010) 72-104,
  arXiv:0912.1366 [hep-th].

\bibitem{BF}
  A.~Bilal and F.~Ferrari,
   Nucl.\ Phys.\  B {\bf 516} (1998) 175
  [arXiv:hep-th/9706145].
  %%CITATION = NUPHA,B516,175;%%


\bibitem{momon}
A.~Ritz and A.~Vainshtein,
  %``Long range forces and supersymmetric bound states,''
  Nucl.\ Phys.\  B {\bf 617} (2001) 43
  [arXiv:hep-th/0102121].


\bibitem{SW1}
N.~Seiberg and E.~Witten, Nucl. Phys. {\bf B426} (1994) 19, hep-th/9407087.

\bibitem{SW2}
N.~Seiberg and E.~Witten, Nucl. Phys. {\bf B431}, 484  (1994)
[hep-th/9408099].

\bibitem{HO} A.~Hanany and Y.~Oz,
  %``On the Quantum Moduli Space of Vacua of $N=2$ Supersymmetric $SU(N_c)$
  %Gauge Theories,''
  Nucl.\ Phys.\  B {\bf 452} (1995) 283
  [arXiv:hep-th/9505075].

\bibitem{APS}
P.~Argyres, M.~Plesser and N.~Seiberg, Nucl. Phys. {\bf B471} (1996)
159, [arXiv:hep-th/9603042].

\bibitem{GMMM} A.Gorsky, A.Marshakov, A.Mironov and A.Morozov,
Phys.Lett., {\bf B380} (1996) 75-80, hep-th/9603140

\bibitem{Gai}
D.~Gaiotto,
  %``N=2 dualities,''
  arXiv:0904.2715 [hep-th].

\bibitem{EgM}
 T.~Eguchi and K.~Maruyoshi,
  %``Penner Type Matrix Model and Seiberg-Witten Theory,''
  JHEP {\bf 1002} (2010) 022
  [arXiv:0911.4797 [hep-th]].


\bibitem{LMN}
A.Losev, A.Marshakov and N.Nekrasov, in Ian Kogan memorial volume
{\it From fields to strings: circumnavigating theoretical physics},
581-621; hep-th/0302191

\bibitem{BE} H.Bateman and A.Erdelyi, {\sl Higher transcendental functions,}
vol.3, 1955.

\bibitem{MN} A.Marshakov and N.Nekrasov,
%``Extended Seiberg-Witten theory and integrable hierarchy,''
  JHEP {\bf 0701} (2007) 104, hep-th/0612019;\\
  A.Marshakov,
  %``On Microscopic Origin of Integrability in Seiberg-Witten Theory,''
  Theor.Math.Phys.  {\bf 154} (2008) 362,
  arXiv:0706.2857.

\bibitem{KriW}
I.~Krichever,
%``The tau-function of the universal Whitham hierarchy,
%matrix models and topological field theories'',
Commun. Pure. Appl. Math. {\bf 47} (1992) 437 [arXiv:
hep-th/9205110].

\bibitem{AD}
P.~C.~Argyres and M.~R.~Douglas,
%``New phenomena in SU(3) supersymmetric gauge theory,''
Nucl. Phys. {\bf B448}, 93 (1995) [arXiv:hep-th/9505062];
%%CITATION = HEP-TH 9505062;%%
\\
%\bibitem{APSW}
P.~Argyres, M.~Plesser, N.~Seiberg, and E.~Witten,
%``New N=2 Superconformal Field Theories in Four Dimensions''
Nucl. Phys. {\bf B461}, 71 (1996)
[arXiv:hep-th/9511154];
%%CITATION = NUPHA,B461,71;%%
\\
T.~Eguchi, K.~Hori, K.~Ito and S.~K.~Yang,
  %``Study of $N=2$ Superconformal Field Theories in $4$ Dimensions,''
  Nucl.\ Phys.\  B {\bf 471} (1996) 430
  [arXiv:hep-th/9603002].


\bibitem{Dorey}
 N.~Dorey,
  %``The BPS spectra of two-dimensional supersymmetric gauge theories with
  %twisted mass terms,''
  JHEP {\bf 9811} (1998) 005
  [arXiv:hep-th/9806056];
\\
 N.~Dorey, T.~Hollowood and D.~Tong,
  %``The BPS spectra of gauge theories in two and four dimensions,''
  JHEP {\bf 9905} (1999) 006
  [arXiv:hep-th/9902134].

\bibitem{SYdual}
M.~Shifman and A.~Yung,
%{\em Non-Abelian duality and confinement in
%\ntwo supersymmetric QCD,}
arXiv:0904.1035 [hep-th].


\end{thebibliography}
\end{document}